\documentclass{iaus}
\usepackage{graphicx}
\usepackage{natbib}

\newcommand{\mbh}{\,\hbox{$M_{\rm BH}$}}
\newcommand{\msigma}{\,\hbox{$M_{\rm BH}-\sigma$}}

\newcommand{\um}{\,\hbox{$\mu$m}}

\begin{document}

\topmargin 1.2 in
\oddsidemargin 1.5 in
\evensidemargin 1.5 in

\title[Symposium 267] 
{On the black hole mass - velocity dispersion relation for type-1 and type-2 AGN}

\author[K. M. Dasyra et al.]   
{Kalliopi M. Dasyra$^{1,2}$ , Bradley M. Peterson$^3$ , Linda J. Tacconi$^4$ , Hagai Netzer$^5$ , Luis
C. Ho$^6$ , George Helou$^1$, Lee Armus$^1$ , Dieter Lutz$^4$ , Richard Davies$^4$, \and Linda Watson$^3$}

\affiliation{
$^1$  Spitzer Science Center, California Institute of Technology, USA,
$^2$  Service d' Astrophysique, Commissariat \`a l' Energie Atomique, France,
$^3$  Ohio State University, USA,
$^4$  Max Planck Institute for Extraterrestrial Physics, Germany,
$^5$  Wise Observatory, Israel,
$^6$  Carnegie Observatories, USA.
}

\pubyear{2009}
\volume{267}  
\pagerange{?--?}
\setcounter{page}{1}
\jname{Co-Evolution of Central Black Holes and Galaxies}
\editors{B.M.\ Peterson, R.S.\ Somerville, \& T.\ Storchi-Bergmann, eds.}

\maketitle

\begin{abstract}
 We present results from infrared spectroscopic projects that aim to test the relation
between the mass of a black hole, \mbh , and the velocity dispersion of the stars 
in its host-galaxy bulge. We demonstrate that near-infrared, high-resolution spectroscopy 
assisted by adaptive optics is key in populating the high-luminosity end of the relation. 
We show that the velocity dispersions of mid-infrared, high-ionization lines originating 
from gas in the narrow-line region of the active galactic nucleus follow the same 
relation. This result provides a way of inferring M$_{BH}$ estimates for the cosmologically 
significant population of obscured, type-2 AGN that can be applicable to data from 
spectrographs on the next generation infrared telescopes.


\end{abstract}

\firstsection 

\section{Introduction}

The comparison of the star-formation history of galaxies with the accretion rate of black holes
(BHs) at different redshifts is of particular importance for our understanding of their evolution (e.g., 
\citealt{marconi04}; \citealt{netzer09}). A challenge in this comparison is the determination of 
BH masses,  \mbh , in intermediate and high-$z$ galaxies, as well as in obscured active galactic 
nuclei (AGN).

For nearby sources, \mbh\ can be inferred from stellar orbits (e.g., \citealt{kormendy98}; 
\citealt{genzel00}; \citealt{davies06}), gas or maser kinematics (e.g., \citealt{miyoshi95}), and reverberation 
experiments (\citealt{peterson04}). Most of these techniques often fail for sources at  $z$$>$0.5  
due to angular resolution and sensitivity limitations of the existing instrumentation. At such redshifts,
\mbh\ is often estimated indirectly, using the relation between the mass of a BH and the velocity 
dispersion $\sigma$ of the stars in its galaxy bulge (\citealt{ferrarese00}; \citealt{gebhardt00}; 
\citealt{gultekin}).  The stellar $\sigma$ value is typically substituted by the velocity dispersion of 
the narrow-line-region (NLR) gas clouds, which is measured from the 5007\AA\  [O III] line assuming that 
the NLR kinematics are primarily determined by the gravitational potential of the bulge (\citealt{shields03}; 
\citealt{greene05}).

However, the obscuration of the optical [O III] line can be non negligible (\citealt{kauffmann03}). This can 
be a problem for NLR kinematic studies of obscured, type-2 sources, which constitute a cosmologically 
significant AGN population (e.g., \citealt{gilli07}; \citealt{lacy07}). We aimed to extend this technique in 
the MIR,  where the NLR lines suffer little from obscuration. Moreover, the [Ne V] and [O IV] lines have a 
higher ionization potential than [O III] and are therefore likely to originate from clouds that are closer to the 
BH. 

\section{The existence of an \msigma\ relation for the NLR gas emitting in the MIR.}

\begin{figure}[!h]
\begin{center}
 \includegraphics[width=3.0in,height=2.3in]{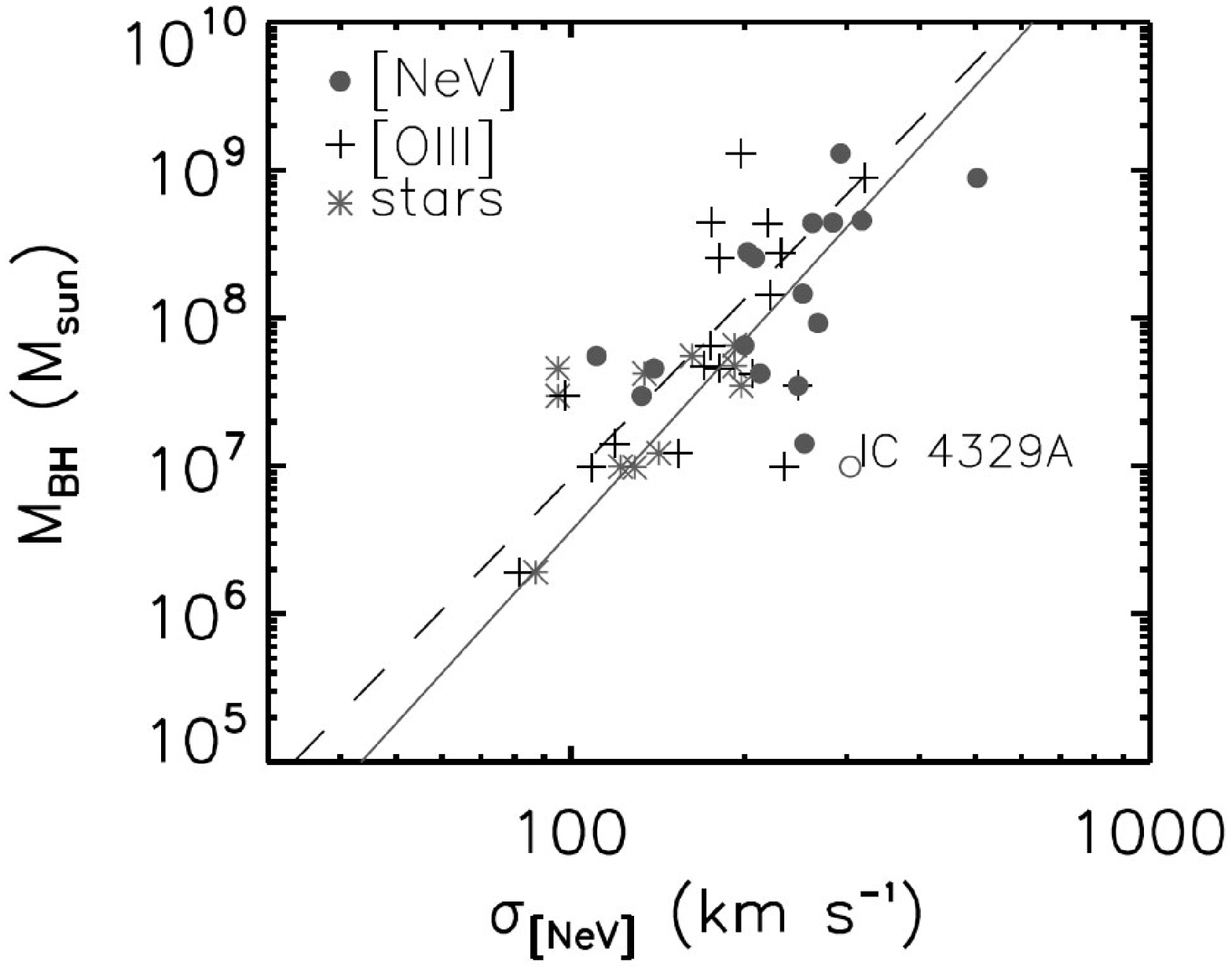} 
 \caption{The relation between \mbh\ and the velocity dispersion of a NLR emission line in the MIR, [Ne V],  
 for reverberation-mapped AGN (\citealt{dasyra08}). The solid line corresponds to the best-fit solution to the [Ne V] 
 velocity dispersion values (excluding IC4329A) and the  dashed line corresponds to the \cite{tremaine02} 
 relation. The stellar velocity dispersion measurements of the same sources are plotted as stars (\citealt{onken04};
\citealt{nelson04}). Similar results are found for the [O IV] line width (\citealt{dasyra08}).}
   \label{fig:nev}
\end{center}
\end{figure}

We analyzed {\it Spitzer} IRS and ISO SWS archival data of local AGN with \mbh\ measurements from reverberation 
experiments (\citealt{peterson04}). We detected resolved [Ne V] and [O IV]  lines at 14.32 \um\ and 25.89 \um , 
respectively,  in more than half of the sources in our sample (\citealt{dasyra08}). The resolution-corrected $\sigma$ 
measurements of both [Ne V] and [O IV] follow impressively well the \msigma\ relation (\citealt{dasyra08}; 
see Fig.\,\ref{fig:nev} for [Ne V]) with a scatter that is comparable to that of the stellar velocity dispersions 
(\citealt{onken04}; \citealt{nelson04}), supporting previous findings that the NLR gas is often gravitationally 
bound to the bulge.  

This result can have various applications for high-resolution MIR spectra from future IR observatories such 
as JWST and SPICA. Since the \msigma\ relation holds for the NLR gas emitting in the MIR, it can be applied to 
distant and obscured sources with resolved MIR narrow lines to derive estimates of their BH masses. By comparing BH 
masses of type-1 and type-2 AGN at similar redshifts, it can also provide a testbed for the AGN unification model.

The flux of [Ne V] or [O IV] also correlates well with the optical 5100 \AA\ continuum flux (\citealt{schweitzer06}; 
\citealt{dasyra08}) for the observed AGN, indicating that it can be used as a proxy of the bolometric 
AGN luminosity assuming standard bolometric conversion factors (\citealt{elvis94}; \citealt{marconi04}).
Therefore, high-resolution MIR spectroscopy can also provide a means to derive the Eddington accretion 
rates for distant and obscured AGN.

\section{A way to improve the calibration of the local stellar \msigma\ relation.}

In order to determine the accretion rate of BHs at intermediate and high $z$ using gas kinematics 
and the  \msigma\  relation, it is necessary to test whether the relation remains identical with look-back 
time. Various studies in the literature suggest that an evolution of the relation could be possible (e.g., 
\citealt{salviander07}; \citealt{treu07}; \citealt{woo08}). However, determining or quantifying such an 
evolution remains highly uncertain. The limitations are primarily originating from the selection 
biases of distant sources, and from the lack of a robust calibration of the AGN \msigma\ relation in 
high-luminosity AGN in the local Universe.  

\begin{figure}[!h]
\begin{center}
 \includegraphics[width=2.5in,height=1.8in]{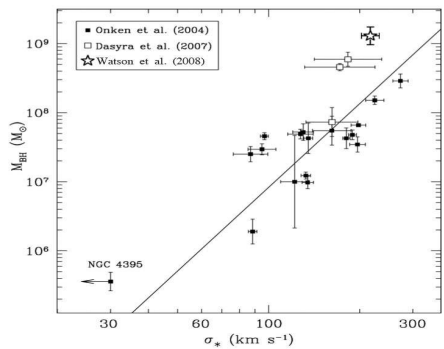} 
 \caption{ The AGN \mbh\ sigma relation including the host galaxies of some reverberation-mapped
 QSOs. Adapted from \cite{watson08}. }
   \label{fig:ms}
\end{center}
\end{figure}

The sources that were initially used for the construction of the local AGN relation were primarily 
low-luminosity AGN. In QSOs, the absorption features that are traditionally 
used for the extraction of the stellar $\sigma$, such as the Ca II triplet at 8498, 8542, and 8662\AA\ 
were undetected, heavily diluted by the bright AGN continuum. We demonstrated that in such cases
the stellar $\sigma$ can be measured from the $H$-band CO bandheads (\citealt{dasyra07}). The 
advantage of using NIR instead of optical lines is that the AGN spectral energy distribution has a local 
minimum (\citealt{elvis94}) and the host galaxy has a local maximum close to 1.6 \um . The QSOs with 
\mbh\ and $\sigma$ measurements  were located above the high-mass end of the relation (\citealt{dasyra07}; 
\citealt{watson08};  Fig.\,\ref{fig:ms}). Differences in the kinematics of stellar populations alone could 
not account for the observed discrepancy.

It is possible that the calibration of the BH masses in QSOs requires modifications. Specifically, 
the statistically-determined value of the factor $f$ that converts the virial mass enclosed in the broad 
line region to the BH mass (\citealt{onken04}) could require a different prescription for AGN of high- 
and low- accretion rates (e.g. \citealt{collin06}).  Alternatively, a steepening of the slope of the relation 
at the high luminosity or mass end is plausible, as it has been suggested based on [O III] kinematics 
(\citealt{gaskell09}). Given the uncertainties in the slope of the local relation, its redshift evolution 
should be examined with caution.

\begin{figure}[!h]
\begin{center}
 \includegraphics[width=5.5in]{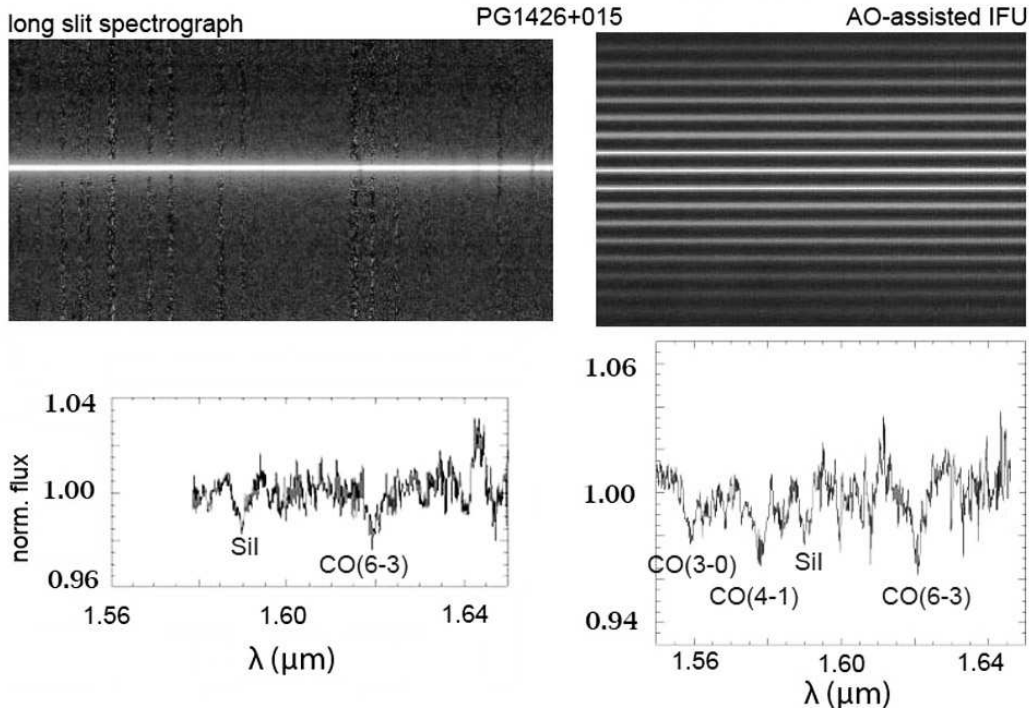} 
 \caption{ $H$-band spectra of PG1426+015, obtained with the long-slit spectrograph ISAAC at 
 the VLT (left panels; data from \citealt{dasyra07}) and the IFU NIFS on GEMINI north (right panels;
data used in \citealt{watson08}). The stellar CO  absorption features are more prominent in the IFU 
data despite the  $\sim$2 times shorter on-source integration time. }
   \label{fig:ifu}
\end{center}
\end{figure}

To further study the behaviour of the high-luminosity end of the relation,  we have acquired NIR integral
field unit (IFU) datasets with SINFONI and NIFS (Grier et al. in preparation) that are assisted by adaptive 
optics.  An example, PG1426+015, was presented in \cite{watson08},  which shows a significant 
improvement in S/N ratio between the long-slit and the IFU spectra (Fig.\,\ref{fig:ifu}). The success 
of the IFU data relies in spatially disentangling the AGN from the host galaxy spectra (Fig.\,\ref{fig:ifu}).

\section{Conclusions}

The combination of {\it Spitzer} IRS and ISO SWS spectroscopy enabled the study of NLR gas kinematics 
in the MIR for reverberation-mapped AGN. The stellar \msigma\ relation holds within the errors for NLR 
gas clouds that emit in the MIR as determined by the [Ne V] and [O IV] emission line widths. Therefore, 
resolved MIR narrow lines (in combination with the \msigma\ relation) can be used to estimate the 
masses of BHs in obscured AGN. The measurement of stellar kinematics of QSO 
host galaxies can be achieved with NIR medium-to-high resolution long-slit and IFU spectroscopy.
The hosts of local high-luminosity AGN have higher BH masses than those that the \msigma\ relation would 
predict for their velocity dispersions, similar to what has been found for high-$z$ QSOs. This dictates the need 
for a better calibration of the relation prior to studying its possible evolution with $z$.

\end{document}